# Coupled quantum oscillators within independent quantum reservoirs

Illarion Dorofeyev[*]

Institute for Physics of Microstructures, Russian Academy of Sciences,

603950, GSP-105 Nizhny Novgorod, Russia

**Abstract**

System of the quantum Langevin equations for two quantum coupling oscillators within independent heat baths of quantum oscillators are obtained using a model Hamiltonian and corresponding Heisenberg equations of motion. Expressions for mean energy of coupled oscillators and their mean energy of interaction are derived and analyzed. Nonmonotonic dependence of the interaction energy versus a coupling constant is demonstrated and explained. Nontrivial dependence of the quantities as a consequence of the difference in temperatures of heat baths is shown.

**Corresponding author**: Illarion Dorofeyev

Institute for Physics of Microstructures Russian Academy of Sciences

603950 Nyzhny Novgorod, GSP-105, Russia

Phone:+7 910 7934517

E-mail1: dorof@ipm.sci-nnov.ru

E-mail2: Illarion1955@mail.ru

A quantum oscillator coupled to a heat bath is a very important and useful problem for many processes dealing with dynamics of open quantum systems. Brownian dynamics of a dissipative quantum system is often modeled by coupling the system to a large number of harmonic oscillators. Such a pure mechanical hamiltonian system composed by one particle plus a thermal reservoir allows investigating the origin of irreversibility in the dynamics of a quantum system interacting with a heat bath. A total Hamiltonian in, various approximations and associated quantum Langevin equations are described in detail in many papers in different contexts; see for example [1-7] and cited literature therein. The key point of the problem is how to model the bath. In this connection the frequently used independent-oscillator model of a heat bath, the velocity coupling model, the linear coupling model and their connections via some unitary transformations were discussed. The impressive result of the theory is a general self-consistent representation of the fluctuation and dissipation interaction of the selected oscillator with a reservoir, analytically expressed by the quantum Langevin equation. The adjoint problem is associated with defining and calculating the partition function, density of states, energy, specific heat of the damped quantum oscillator. In this connection, the corresponding thermodynamic functions for an oscillator coupled to a heat bath, various definitions of the functions, some subtleties due to possible approximations were studied in [8-12]. For instance, an analytic formula for the free energy of a quantum oscillator coupled to a heat bath expressed via the Green's function of the quantum Lagevin equation was derived by two independent methods. Besides, specific heat anomalies of the same open system was investigated based on different possible definitions of the specific heat. Actually in parallel, another set of studies were performed from the point of view of the relaxation of open systems to the equilibrium state and description of the nonequilibrium transport phenomena. A problem for coupled oscillators interacting with heat baths characterized by its own temperatures was considered in [13-18] to clarify the approach to the equilibrium or to some transient stationary state of the system. It was shown that an arbitrary initial state of a harmonic oscillator state decayes towards a uniquely determined stationary state. Exact calculations of heat flows and thermal conductivities in harmonic oscillator chain were made. The above mentioned problem is similar to the more realistic problem of interaction between systems within different thermostats kept with own temperatures, for instance, in physics of the dispersion forces between bodies out of thermal equilibrium [19-25]. Obviously, that the problem of interaction of particles in nonequilibrium systems allows better understanding of natural phenomena.

In this letter the problem of two coupled quantum oscillators interacting with ensembles of harmonic oscillators is considered. These ensembles simulate heat baths characterized, in general, by different temperatures. The main goal of the paper is to derive the interaction energy of the selected two damped oscillators in a system out of thermal equilibrium.

Let us consider two coupled quantum particles surrounded by a large number of heat baths oscillators. The Hamiltonian of the system can be written as follows

$$\hat{H} = \hat{p}_1^2/2m_1 + \hat{V}_1(\hat{x}) + \hat{p}_2^2/2m_2 + \hat{V}_2(\hat{y}) - \lambda\left(\hat{x}\hat{y} + \hat{y}\hat{x}\right)/2 +$$
$$+\sum_{j=1}^{N_1}\left[\hat{p}_j^2/2m_j + m_j\omega_j^2(\hat{q}_j - \hat{x})^2/2\right] + \sum_{k=1}^{N_2}\left[\hat{p}_k^2/2m_k + m_k\omega_k^2(\hat{q}_k - \hat{y})^2/2\right], \quad (1)$$

where $\hat{x}$, $\hat{y}$, $\hat{p}_{1,2}$, $m_{1,2}$ are the particles coordinates and momentum operators and masses, $\hat{V}_1(\hat{x})$, $\hat{V}_2(\hat{y})$ are the potential energies of external forces, $\lambda$ is the coupling constant, $\hat{q}_j, \hat{p}_j, \omega_j, m_j$ and $\hat{q}_k, \hat{p}_k, \omega_k, m_k$ are the oscillators coordinates and momentum operators composing the baths and their eigenfrequencies and masses. The operators are subjected by usual commutation rules $[\hat{x}, \hat{p}_1] = i\hbar$, $[\hat{y}, \hat{p}_2] = i\hbar$, $[\hat{q}_i, \hat{p}_j] = i\hbar \delta_{ij}$ for both baths. In this work we take that the commutators of operators from different reservoirs vanish. It is follows from Eq.(1) that the total Hamiltonian can be represented as a sum of separate terms

$$\hat{H} = \hat{H}_{01} + \hat{H}_{02} + \hat{H}_{int} + \hat{H}_{B1} + \hat{H}_{B2}, \qquad (2)$$

where each term has an obvious meaning after a term-by-term comparison of Eq.(1) and (2).

It should be noted that the Hamiltonian in Eq.(1) is direct extension of the independent oscillator model [4] to the described case of two heat baths with explicit term related to the interaction energy $\hat{H}_{int} = -\lambda(\hat{x}\hat{y} + \hat{y}\hat{x})/2$. Besides, being written by creation-annihilation operators, the Hamiltonian in Eq.(1) looks similar to the one in [16]. But, we have two essential differences. Firstly, the authors of [16] neglected indirect interaction of the selected oscillators through the heat bath. Opposite to this, we include the indirect interaction manifesting in functional dependencies of the operators $\hat{x} = \hat{x}(\hat{q}_j, \hat{p}_j, \hat{q}_k, \hat{p}_k)$ and $\hat{y} = \hat{y}(\hat{q}_j, \hat{p}_j, \hat{q}_k, \hat{p}_k)$ on the bath operators. In this case the operators become non-commuting and the interaction term $\hat{H}_{int}$ should be taken in the symmetrized form in order to obtain real average values. This allows taking into account any weak interactions between the oscillators. Secondly, in our problem the fluctuating forces appear in corresponding quantum motion equations by the self-consistent manner, different from the cited paper [16].

Then we omit the carets for convenience. Use of the Heisenberg equations of motion as applied to the operators $x, p_1$, $y, p_2$ and $q_j, p_j$, $q_k, p_k$, followed by elimination of the operators $p_1, p_2$ and $q_j, p_j$, $q_k, p_k$, yields to the coupled system of quantum Langevin equations

$$\begin{cases} \ddot{x}(t) + \int_{-\infty}^{t} dt' \mu_1(t-t') \dot{x}(t') + V_1'(x)/m_1 = (\lambda/m_1)y + F_1(t), \\ \ddot{y}(t) + \int_{-\infty}^{t} dt' \mu_2(t-t') \dot{y}(t') + V_2'(x)/m_2 = (\lambda/m_2)x + F_2(t), \end{cases} \qquad (3)$$

where $F_{1,2}(t)$ are the operator-valued random forces acting wthin two thermal baths

$$F_1(t) = \sum_{j=1}^{N_1} (m_j / m_1)\omega_j^2 \left[ q_j Cos(\omega_j t) + (p_j / m_j \omega_j) Sin(\omega_j t) \right],$$

$$F_2(t) = \sum_{k=1}^{N_2} (m_k / m_2)\omega_k^2 \left[ q_k Cos(\omega_k t) + (p_k / m_k \omega_k) Sin(\omega_k t) \right] \quad (4)$$

$\mu_{1,2}(t)$ are the memory functions, see, for instance, [4,7].

Then, we use the following Fourier transform to solve the system of Eq.(3)

$$A(t) = \int_{-\infty}^{\infty} d\Omega\, A(\Omega) \exp(-i\Omega t), \quad (5)$$

and consider harmonic potentials $V_1 = m_1 \omega_{01}^2 x^2 / 2$ and $V_2 = m_2 \omega_{02}^2 y^2 / 2$ in Eq.(3), where $\omega_{01}$ and $\omega_{02}$ are the eigenfrequencies of two selected oscillators. Thus, we have

$$\begin{cases} \beta_1(\Omega) x(\Omega) = (\lambda / m_1) y(\Omega) + F_1(\Omega), \\ \beta_2(\Omega) y(\Omega) = (\lambda / m_2) x(\Omega) + F_2(\Omega), \end{cases} \quad (6)$$

instead of the system in Eq.(3), where $\beta_m(\Omega) = (\omega_{0m}^2 - \Omega^2) - i\Omega \mu_{1,2}(\Omega)$, $(m = 1, 2)$, and $\mu_{1,2}(\Omega)$ are the one-side Fourier transforms of the memory functions. From Eqs.(6) we have

$$x(\Omega) = \frac{(\lambda / m_1) F_2(\Omega) + \beta_2(\Omega) F_1(\Omega)}{D(\Omega)}, \quad y(\Omega) = \frac{(\lambda / m_2) F_1(\Omega) + \beta_1(\Omega) F_2(\Omega)}{D(\Omega)}, \quad (7)$$

where $D(\Omega) = \beta_1(\Omega)\beta_2(\Omega) - \lambda^2 / m_1 m_2$.

Now, judging by Eq. (7), one can conclude that the Brownian vibrations $x(\Omega)$, $y(\Omega)$ of the first and second oscillators are determined by the random forces $F_1(\Omega)$, $F_2(\Omega)$ both of the first and second reservoirs due to the constant connection $\lambda$, even in the case of independent heat baths. The Fourier transforms of $F_1(\Omega)$, $F_2(\Omega)$ can be easily found from Eq. (4)

$$F_1(\Omega) = \sum_{j=1}^{N_1} (m_j / m_1)\omega_j^2 \big\{ q_j [\delta(\Omega - \omega_j) + \delta(\Omega + \omega_j)] / 2 + ip_j [\delta(\Omega - \omega_j) - \delta(\Omega + \omega_j)] / 2 m_j \omega_j \big\} \quad (8)$$

$$F_2(\Omega) = \sum_{k=1}^{N_2} (m_k / m_2)\omega_k^2 \big\{ q_k [\delta(\Omega - \omega_k) + \delta(\Omega + \omega_k)] / 2 + ip_k [\delta(\Omega - \omega_k) - \delta(\Omega + \omega_k)] / 2 m_k \omega_k \big\}. \quad (9)$$

From Eq. (7) it is follows that the position fluctuations of two coupled oscillators can be separated in two parts $x(\Omega) = x_1(\Omega) + x_2(\Omega)$ and $y(\Omega) = y_1(\Omega) + y_2(\Omega)$ due to different bath actions, where

$$x_1(\Omega) = \frac{(\lambda/m_1)F_2(\Omega)}{D(\Omega)}, \quad x_2(\Omega) = \frac{\beta_2(\Omega)F_1(\Omega)}{D(\Omega)}, \tag{10}$$

$$y_1(\Omega) = \frac{(\lambda/m_2)F_1(\Omega)}{D(\Omega)}, \quad y_2(\Omega) = \frac{\beta_1(\Omega)F_2(\Omega)}{D(\Omega)}. \tag{11}$$

Using Eq. (5) we obtain $x_1(t)$, $x_2(t)$ and $y_1(t)$, $y_2(t)$.

From the Hamiltonian in Eqs. (1),(2) we calculate mean values $\langle H_{01} \rangle$, $\langle H_{02} \rangle$ and $\langle H_{int} \rangle$. In this connection it is worth to recall that in the Ohmic approach the quantities $\langle H_{01} \rangle$ and $\langle H_{02} \rangle$ coincide with the mean energies of oscillators [8]. Thus, we consider the nonmarkovian case, when $\mu_{1,2}(\Omega) = 2\gamma_{1,2}$. Direct calculations of $x^2(t)$ and $y^2(t)$ followed by a time and ensemble averaging using the mean values for the baths variables $\langle q_n q_m \rangle = (\hbar/2m_n\omega_n)Coth(\hbar\omega_n/2k_BT)\delta_{nm}$, $\langle p_n p_m \rangle = (\hbar m_n\omega_n/2)Coth(\hbar\omega_n/2k_BT)\delta_{nm}$, $\langle q_n p_m \rangle = -\langle p_n q_m \rangle = (1/2)i\hbar\delta_{nm}$ from papers [4], [26], in case of independent thermal baths yields for the mean energy of the first oscillator

$$\bar{U}_1(T_1,T_2) = m_1 \langle \dot{x}^2(t) \rangle/2 + m_1\omega_{01}^2 \langle x^2(t) \rangle/2 =$$
$$\frac{1}{2}\left\{ \frac{\lambda^2}{m_2^2}\sum_{k=1}^{N_2}\left(\frac{m_k}{m_1}\right)\omega_k^2 \frac{(\omega_k^2+\omega_{01}^2)\Theta(\omega_k,T_2)}{|D(\omega_k)|^2} + \sum_{j=1}^{N_1}\left(\frac{m_j}{m_1}\right)\omega_j^2 \frac{(\omega_j^2+\omega_{01}^2)\Theta(\omega_j,T_1)|\beta_2(\omega_j)|^2}{|D(\omega_j)|^2} \right\}, \tag{12}$$

for the mean energy of the second oscillator

$$\bar{U}_2(T_1,T_2) = m_2 \langle \dot{y}^2(t) \rangle/2 + m_2\omega_{02}^2 \langle y^2(t) \rangle/2 =$$
$$\frac{1}{2}\left\{ \frac{\lambda^2}{m_1^2}\sum_{j=1}^{N_1}\left(\frac{m_j}{m_2}\right)\omega_j^2 \frac{(\omega_j^2+\omega_{02}^2)\Theta(\omega_j,T_1)}{|D(\omega_j)|^2} + \sum_{k=1}^{N_2}\left(\frac{m_k}{m_2}\right)\omega_k^2 \frac{(\omega_k^2+\omega_{02}^2)\Theta(\omega_k,T_2)|\beta_1(\omega_k)|^2}{|D(\omega_k)|^2} \right\}, \tag{13}$$

and for the mean energy of their interaction

$$\bar{U}_{int}(T_1,T_2) = -\lambda\left[\langle x(t)y(t) \rangle + \langle y(t)x(t) \rangle\right]/2 =$$
$$-\frac{\lambda^2}{m_1 m_2}\left\{\sum_{j=1}^{N_1}\left(\frac{m_j}{m_1}\right)\omega_j^2 \frac{\Theta(\omega_j,T_1)\operatorname{Re}\{\beta_2(\omega_j)\}}{|D(\omega_j)|^2} + \sum_{k=1}^{N_2}\left(\frac{m_k}{m_2}\right)\omega_k^2 \frac{\Theta(\omega_k,T_2)\operatorname{Re}\{\beta_1(\omega_k)\}}{|D(\omega_k)|^2}\right\}, \tag{14}$$

where $\Theta(\omega_n,T) = (\hbar\omega_n/2)Coth(\hbar\omega_n/2k_BT)$ is the mean energy of free oscillator within the baths. Transition to the quasicontinuum spectrum is performed as usual by introducing the spectral density $\eta(\omega)$. Thus, introducing the spectral densities for two baths as follows

$$\rho_{1,2}(\omega) = \eta_{1,2}(\omega)\pi m(\omega)\omega^2/2m_{1,2}, \tag{15}$$

we obtain from Eqs.(12)-(14) the corresponding formulas in case of the quasicontinuum spectral distribution of oscillators. For the mean energy of the first oscillator we have

$$\bar{U}_1(T_1,T_2) = \frac{\lambda^2}{m_2^2}\int_0^{\Omega_{max}} \frac{d\omega}{\pi}\rho_2(\omega)\frac{(\omega^2+\omega_{01}^2)\Theta(\omega,T_2)}{|D(\omega)|^2}$$
$$+\int_0^{\Omega_{max}}\frac{d\omega}{\pi}\rho_1(\omega)\frac{(\omega^2+\omega_{01}^2)\Theta(\omega,T_1)|\beta_2(\omega)|^2}{|D(\omega)|^2},$$
(16)

for the mean energy of the second oscillator

$$\bar{U}_2(T_1,T_2) = \int_0^{\Omega_{max}}\frac{d\omega}{\pi}\rho_2(\omega)\frac{(\omega^2+\omega_{02}^2)\Theta(\omega,T_2)|\beta_1(\omega)|^2}{|D(\omega)|^2}$$
$$+\frac{\lambda^2}{m_1^2}\int_0^{\Omega_{max}}\frac{d\omega}{\pi}\rho_1(\omega)\frac{(\omega^2+\omega_{02}^2)\Theta(\omega,T_1)}{|D(\omega)|^2},$$
(17)

and for the mean energy of their interaction

$$\bar{U}_{int}(T_1,T_2) = -\frac{2\lambda^2}{m_1 m_2}\left\{\int_0^{\Omega_{max}}\frac{d\omega}{\pi}\rho_1(\omega)\frac{\Theta(\omega,T_1)\text{Re}\{\beta_2(\omega)\}}{|D(\omega)|^2}\right.$$
$$\left.+\int_0^{\Omega_{max}}\frac{d\omega}{\pi}\rho_2(\omega)\frac{\Theta(\omega,T_2)\text{Re}\{\beta_1(\omega)\}}{|D(\omega)|^2}\right\} = \int_0^{\Omega_{max}}d\omega\,\bar{U}_{int}(\omega),$$
(18)

where in the second equality in Eq. (18) we introduced the spectral power density $\bar{U}_{int}(\omega)$ of the mean interaction energy.

It is easy to verify that for decoupled identical oscillators ($\lambda=0$, $\omega_{01,02}=\omega_0$, $m_{1,2}=m$) in case of total equilibrium ($T_1=T_2=T$) from Eqs. (16)-(18) one can obtain the well known quantity of mean energy of an oscillator within a thermal bath

$$U(T) = \int_0^{\Omega_{max}}\frac{d\omega}{\pi}2\gamma\frac{(\omega^2+\omega_0^2)\Theta(\omega,T)}{(\omega^2-\omega_0^2)^2+4\gamma^2\omega^2},$$
(19)

where we have taken into account Eq.(15) and the relation $\pi\eta(\omega_0)|g(\omega_0)|^2 \approx \gamma$ from [27] and $g(\omega_0)\approx\sqrt{\omega_0^2 m(\omega_0)/m}$, which can be derived in the rotation wave approximation.

Numerical calculations were performed using the Debye and Gauss spectral densities for $\rho_{1,2}(\omega)$ in Eqs. (16)-(18). Accordingly we chose the expressions $\rho_{1,2}(\omega) = 2\gamma(\omega/\omega_{01,2})^2$ and $\rho_{1,2}(\omega) = 2\gamma\exp[-(\omega-\omega_{01,2})^2/2\sigma^2]$ in order to obtain the limiting case $\rho_{1,2}(\omega)\to 2\gamma$ at $\omega\to\omega_{01,2}$. Integration in Eqs. (16)-(18) was performed up to the Debye frequency ( $\Omega_{max} = \omega_D \approx 10 \div 50 \omega_{01,2}$) and up to infinity for the Debye and Gauss models, correspondingly. The chosen parameters of oscillators are as follows: $m_1 = m_2 = 10 m_a = 10^{-23} g$, where $m_a$ is the atomic unit mass, $\omega_{01} = 10^{13} rad \cdot s^{-1}$, typically for the infrared spectra of vibrating atoms in solids.

First of all we would like to show some features in the spectral composition of the interaction energy. Figure 1 exemplifies the spectral power density $\bar{U}_{int}(\omega)$ of the interaction energy of two damped oscillators versus the normalized frequency $\omega/\omega_{01}$ at different coupling constants $\lambda/\lambda_0$ shown in the insets. The constants are normalized to the value of the spring constant $\lambda_0 = m_1 \omega_{01}^2 = 1000 g (rad/s)^2$ of the first oscillator. The solid and dotted curves in each picture correspond to the pair of oscillators when the eigenfrequency of the second oscillator is larger ($\omega_{02} = 1.3\omega_{01}$) and smaller ($\omega_{02} = 0.5\omega_{01}$) than the eigenfrequency of the first oscillator. The damping of the oscillators are $\gamma_i = 0.02\omega_{0i}, i = 1,2$. Despite of the given arbitrary units of $\bar{U}_{int}(\omega)$ the relative ratios between the peaks in each picture are true. It is clear from the left panel of the figure that at the weak coupling between the oscillators ($\lambda \ll \lambda_0$) the resonance peaks of $\bar{U}_{int}(\omega)$ are situated exactly at the positions of their eigenfrequencies. Then, the larger the connection constant ($\lambda \leq \lambda_0$), the larger the shifts of the peaks from their undisturbed positions, as it can be seen from the both panels of the figure. At the comparatively strong connection ($\lambda \geq \lambda_0$) the "soft" modes of the two pairs of oscillators shift to the low frequency range ($\omega < \omega_{01}$) up to full disappearance, as it can be seen from the right panel of the figure. The extremely strong connection ($\lambda \gg \lambda_0$) yields eventually the intrusive vibrations of the coupled oscillators at the frequency $\omega^2 \approx \lambda/\sqrt{m_1 m_2}$, which follows from the resonance condition $\text{Re}\{D(\omega)\} = \omega_{01}^2 \omega_{02}^2 + \omega^2(\omega^2 - \omega_{01}^2 - \omega_{02}^2) - \lambda^2/m_1 m_2 = 0$ at $\lambda^2 \gg \omega_{01}^2 \omega_{02}^2 m_1 m_2$. Thus, the obtained formulae in Eq.(18) allows investigating the interaction energy of two oscillators in a wide range of coupling constant.

A nontrivial behavior of the interaction energy $\bar{U}_{int}(T_1, T_2)$ in Eq. (18) versus the coupling constant is demonstrated in Fig.2. Figure (a) exemplifies the normalized quantity $\bar{U}_{int}(T_1, T_2)/[\bar{U}_1(T_1, T_2) + \bar{U}_2(T_1, T_2)]$ as the function of the normalized coupling constant $\lambda/\lambda_0$, where $\bar{U}_{1,2}(T_1, T_2)$ are given by Eqs. (16)-(17). The three curves correspond to three pairs of coupled oscillators: $\omega_{01} = \omega_{02} = 10^{13} rad \cdot \sec^{-1}$ - the curve 1, $\omega_{01} = 10^{13} rad \cdot \sec^{-1}$, $\omega_{02} = 1.1\omega_{01}$ - the curve 2 and $\omega_{01} = 10^{13} rad \cdot \sec^{-1}$, $\omega_{02} = 1.2\omega_{01}$ - the curve 3. The main feature of the plots is the jump of the interaction energy at the coupling values $\lambda \approx \omega_{01}\omega_{02}\sqrt{m_1 m_2}$. It follows directly from the resonance condition $|D(\omega)| \to 0$ at $\omega^2 \ll \omega_{01,2}^2$. Figure (b) shows the nonmonotonic dependence of the interaction energy of the same pairs of oscillators at comparatively small couplings ($\lambda \leq \lambda_0$), which disappears in case of equal temperatures.

Figure 3 demonstrates the normalized interaction energy $\bar{U}_{int}(T_1, T_2) / [\bar{U}_1(T_1, T_2) + \bar{U}_2(T_1, T_2)]$ versus the normalized quantity $\omega_{02}/\omega_{01}$ at two essentially different cases. Namely, this quantity is plotted in figure (a) at $T_1 = T_2 = 0K$ and at $T_1 = 300K$, $T_2 = 1000K$ in figure (b). The numbers near curves correspond to different coupling constants $\lambda = 10^{-2}\lambda_0$ -1, $\lambda = 2 \times 10^{-2}\lambda_0$, and $\lambda = 3 \times 10^{-2}\lambda_0$ -3. It should be emphasized that the two-body dependence in figure (b) is very similar to the dependence of the van der Waals interaction energy in a two-body system out of equilibrium, see, for example Fig.2 and 10 in [24].

In summary, the interaction of two coupled quantum oscillators within independent reservoirs of quantum oscillators is analyzed based on the system of quantum coupled Langevin equations ensuing from the model Hamiltonian. Spectral power density of the interaction energy versus a coupling constant is studied. The nonmonotonic dependence of the interaction energy in a wide range of coupling variation is shown. The nontrivial dependence of the interaction energy as a consequence of the difference in temperatures of heat baths is demonstrated.

**References**

[1] A.O. Caldeira and A. J. Leggett, Phys. Rev. Lett. 46 (1981) 211.

[2] P.S. Riseborough, P. Hänggi, U.Weiss, Phys. Rev. A 31 (1985) 471.

[3] A. J. Leggett *et al.*, Rev. Mod. Phys. 59 (1987) 1.

[4] G.W. Ford, J.T. Lewis and R.F. O'Connell, Phys. Rev. A 37 (1988) 4419.

[5] G.W. Ford, J.T. Lewis and R.F. O'Connell, Ann. Phys. (N.Y.) 185 (1988) 270.

[6] X.L. Li, G.W. Ford and R.F. O'Connell, Am. J. Phys. 61 (1993) 924.

[7] P. Hänggi and G-L. Ingold, Chaos 15 (2005) 026105.

[8] G.W. Ford, J.T. Lewis and R.F. O'Connell, J. Stat. Phys. 53 (1988) 439.

[9] G.W. Ford and R.F. O'Connell, Phys. Rev. B 75 (2007) 134301.

[10] P. Hänggi, G-L Ingold and P. Talkner, New J. Phys. 10 (2008) 115008.

[11] G-L. Ingold, P. Hänggi and P. Talkner, Phys. Rev. E 79 (2009) 061105.

[12] A. Hanke and W. Zwerger, Phys. Rev. E 52 (1995) 6875.

[13] J. Rau, Phys. Rev. 129 (1963) 1880.

[14] M. Bolsterli, M. Rich and W.M. Visscher, Phys. Rev. A 1 (1969) 1086.

[15] M. Rich and W.M. Visscher, Phys. Rev. B 11 (1975) 2164.

[16] R. Glauber and V.I. Man'ko, Zh. Eksp. Teor. Fiz. 87 (1984) 790 [Sov. Phys. JETP 60 (1984) 450].

[17] U. Zürcher and P. Talkner, Phys.Rev. A 42 (1990) 3278.

[18] A. Chimonidou and E.C.G. Sudarshan, Phys. Rev. A 77 (2008) 03212.

[19] I.A. Dorofeyev, J. Phys. A 31 (1998) 4369.

[20] I.A. Dorofeyev, Phys. Lett. A 340 (2005) 251.

[21] M. Antezza, L. P. Pitaevskii, S. Stringari and V. B. Svetovoy, Phys. Rev. Lett. 97 (2006) 223203.

[22] L. P. Pitaevskii, J. Phys. A 39 (2006) 6665.

[23] M. Antezza, L. P. Pitaevskii, S. Stringari and V. B. Svetovoy, Phys. Rev. A 77 (2008) 022901.

[24] I.A. Dorofeyev, Phys. Rev. E 82 (2010) 011602.

[25] R. Messina and M. Antezza, Phys. Rev. A 84 (2011) 042102.

[26] G.W. Ford, M. Kac and P. Mazur, J. Math. Phys. 6 (1965) 504.

[27] L. Mandel and E. Wolf, Quantum coherence and quantum optics, Cambridge University Press, 1995.

**Figure captions**

**FIG. 1.** Spectral power density $\bar{U}_{int}(\omega)$ from Eq. (18) of the interaction energy of two damped oscillators versus the normalized frequency $\omega/\omega_{01}$ at different coupling constants $\lambda/\lambda_0$ shown in the insets. The constants are normalized to the value of the spring constant $\lambda_0 = m_1\omega_{01}^2 = 1000 g(rad/s)^2$ of the first oscillator. The solid and dotted curves in each picture correspond to the pair of oscillators when the eigenfrequency of the second oscillator is larger ($\omega_{02} = 1.3\omega_{01}$) and smaller ($\omega_{02} = 0.5\omega_{01}$) than the eigenfrequency of the first oscillator. The dampings of the oscillators are $\gamma_i = 0.02\omega_{0i}, i=1,2$, the temperatures of heat baths are $T_1 = 300K$ and $T_1 = 700K$, correspondingly. The vertical thin lines indicate the positions of eigenfrequencies of two pairs of free oscillators at $\omega = \omega_{01}$, $\omega = \omega_{02} = 1.3\omega_{01}$ and at $\omega = \omega_{01}$, $\omega = \omega_{02} = 0.5\omega_{01}$.

**FIG. 2.** Normalized quantity $\bar{U}_{int}(T_1,T_2)/[\bar{U}_1(T_1,T_2)+\bar{U}_2(T_1,T_2)]$ versus the normalized coupling constant $\lambda/\lambda_0$, where $\bar{U}_{1,2}(T_1,T_2)$ and $\bar{U}_{int}(T_1,T_2)$ are given by Eqs. (16)-(17) and (18) correspondingly. The three curves in figure (a) correspond to three pairs of coupled oscillators: $\omega_{01} = \omega_{02} = 10^{13} rad \cdot sec^{-1}$ - the curve 1, $\omega_{01} = 10^{13} rad \cdot sec^{-1}$, $\omega_{02} = 1.1\omega_{01}$ - the curve 2 and $\omega_{01} = 10^{13} rad \cdot sec^{-1}$, $\omega_{02} = 1.2\omega_{01}$ - the curve 3. The dampings of the oscillators are $\gamma_i = 0.01\omega_{0i}, i=1,2$, the temperatures of heat baths are $T_1 = 300K$ and $T_1 = 1000K$. Figure (b) shows the same quantity at comparatively small couplings ($\lambda \leq \lambda_0$) in more detail.

**FIG.3.** Normalized interaction energy $\bar{U}_{int}(T_1,T_2)/[\bar{U}_1(T_1,T_2)+\bar{U}_2(T_1,T_2)]$ versus the normalized quantity $\omega_{02}/\omega_{01}$ at two essentially different pairs of temperatures. Namely, in figure (a) at $T_1 = T_2 = 0K$ and at $T_1 = 300K$, $T_2 = 1000K$ in figure (b). The numbers near curves correspond to different coupling constants $\lambda = 10^{-2}\lambda_0$ - the curve 1, $\lambda = 2\times 10^{-2}\lambda_0$ - the curve 2 and $\lambda = 3\times 10^{-2}\lambda_0$ - the curve 3.

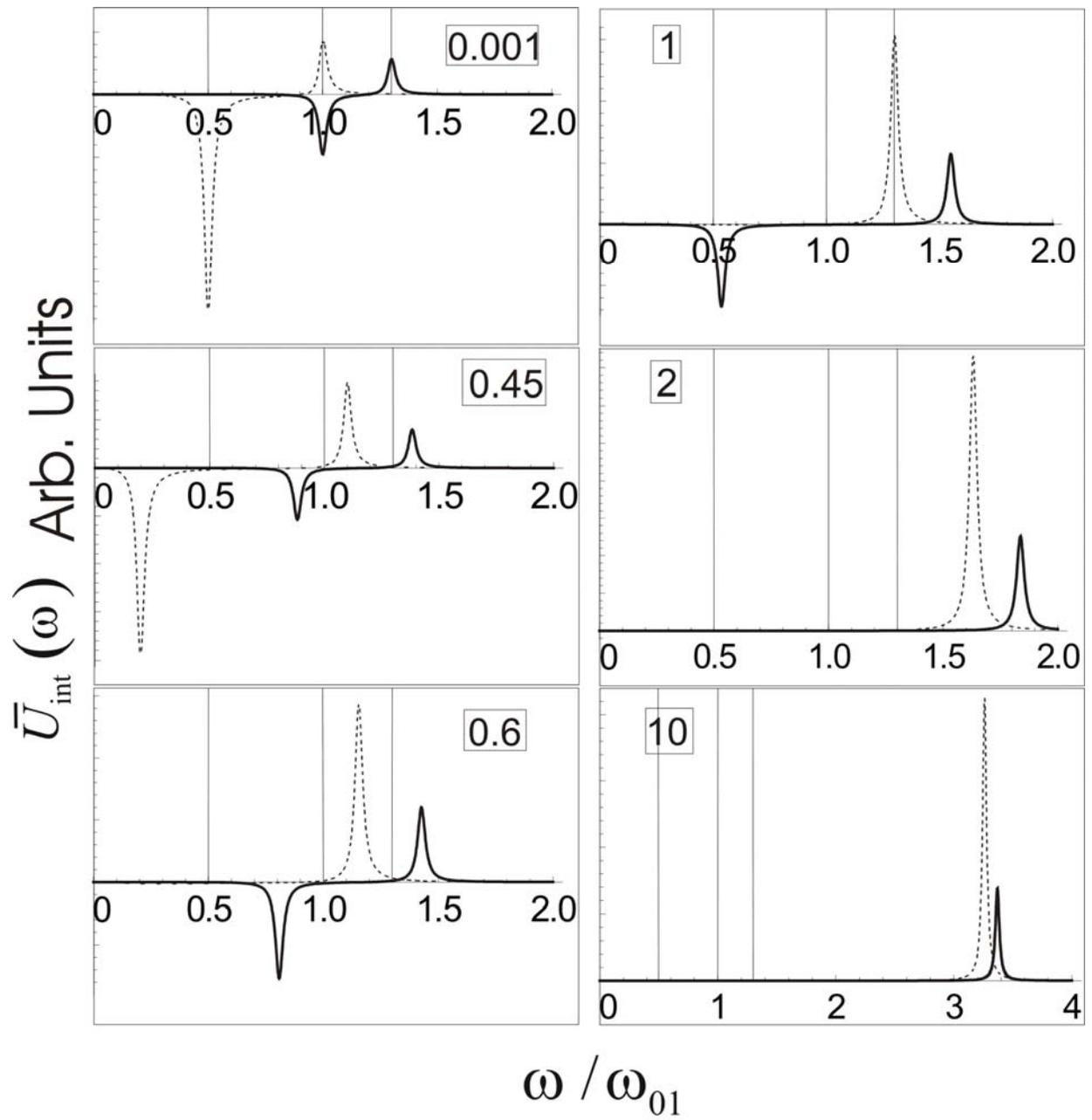

Fig.1

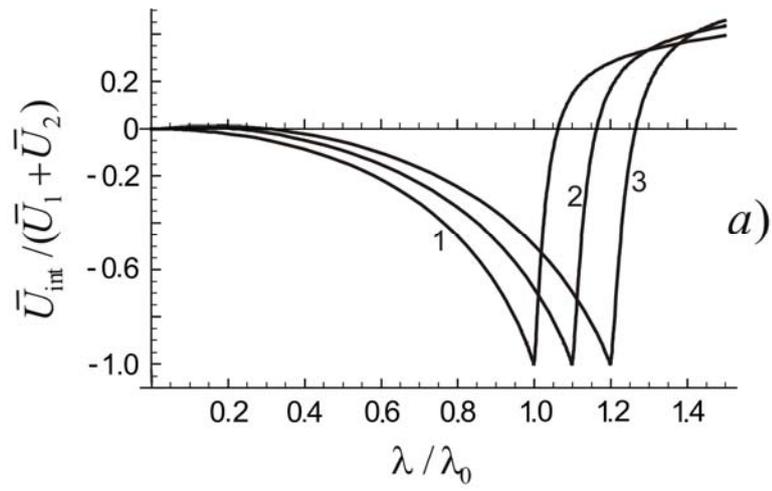

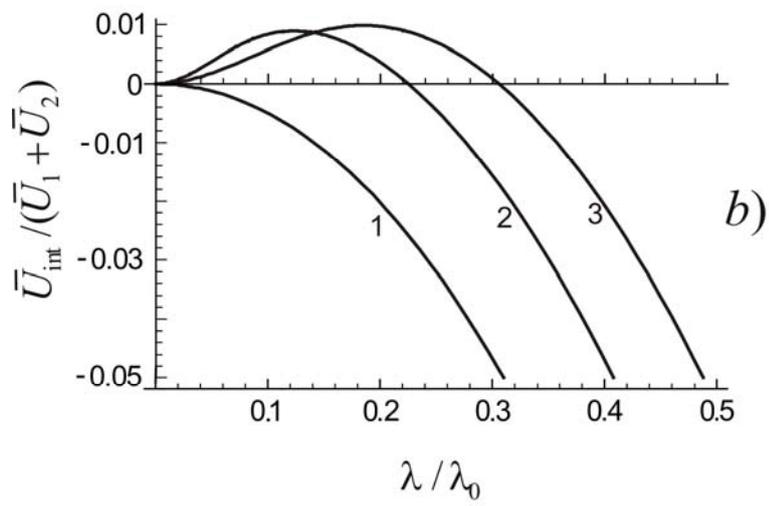

Fig.2

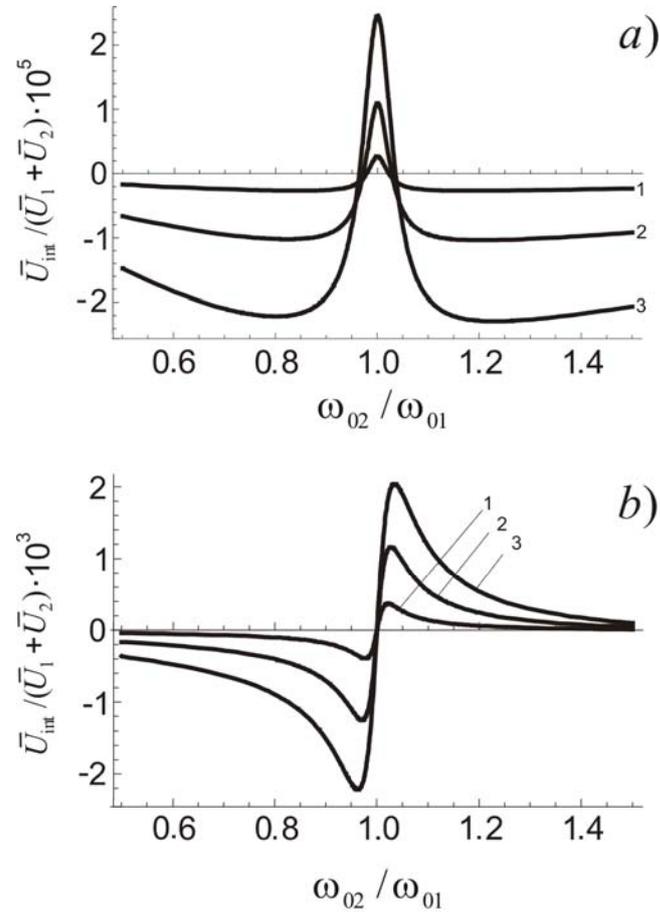

Fig.3